\documentclass[envcountsame,11pt,orivec]{llncs}
\usepackage{a4wide}
\usepackage{xspace}
\usepackage{amsmath}
\usepackage{amssymb}
\usepackage{hyperref,breakurl}
\usepackage{graphicx}
\usepackage{fancyvrb}
\usepackage{footmisc}
\newcommand{\COMMENTED}[1]{}

\newcommand{\lab}[1]{\label{#1}}%\newcommand{\lab}[1]{\label{#1}\comment{#1}}
\newcommand{\myurl}[1]{\href{#1}{\texttt{#1}}\xspace}

\newcommand{\doi}[1]{\href{http://dx.doi.org/#1}{\texttt{doi:#1}}}

\newcommand{\IR}{\mathbb{R}}

\newcommand{\IZ}{\mathbb{Z}}

\newcommand{\IN}{\mathbb{N}}

\newcommand{\dom}{\operatorname{dom}}

\newcommand{\Card}{\operatorname{Card}}

\newcommand{\calLogspace}{\text{\rm\textsf{L}}\xspace}
\newcommand{\calNC}{\text{\rm\textsf{NC}}\xspace}
\newcommand{\calP}{\text{\rm\textsf{P}}\xspace}
\newcommand{\calNP}{\text{\rm\textsf{NP}}\xspace}
\newcommand{\calFP}{\text{\rm\textsf{FP}}\xspace}
\newcommand{\sharpP}{\text{\rm\textsf{\#P}}\xspace}
\newcommand{\calPSPACE}{\text{\rm\textsf{PSPACE}}\xspace}
\newcommand{\calEXP}{\text{\rm\textsf{EXP}}\xspace}

\newcommand{\mycite}[2]{{\rm\cite[\textsc{#1}]{#2}}}

\newcommand{\MAX}{\operatorname{MAX}}

\newcommand{\toto}{\rightrightarrows}

\newcommand{\Gevrey}{G}

\newcommand{\hexp}{h_{\exp}}

\newcommand{\iRRAM}{\textsf{iRRAM}\xspace}

\newcommand{\NAG}{\textsf{NAG}\xspace}

\newcommand{\dfeq}{:=}

\newcommand{\dffn}{:} % define a function
\newcommand{\pcolon}{\mathpunct{\,:\subseteq}}

\newtheorem{myexample}[theorem]{Example}

\newtheorem{challenge}[theorem]{Challenge}
\begin{document}
%\addtocounter{page}{-1}%
\title{Invitation to Real Complexity Theory:
Algorithmic Foundations to Reliable Numerics with Bit-Costs\thanks{%
Partly supported by the 
Japan Society for the Promotion of Science (JSPS) \emph{Core-to-Core Program},
by the \emph{JSPS Kakenhi} project \texttt{26700001}, 
by the \emph{German Research Foundation} (DFG) project \texttt{Zi\,1009/4-1},
and by the \emph{EU FP7 IRSES} project \texttt{294962}.}}
\author{Akitoshi Kawamura (The University of Tokyo) \and Martin Ziegler (KAIST)}
\institute{}
\titlerunning{Invitation to Real Complexity Theory}
\authorrunning{Akitoshi Kawamura and Martin Ziegler}
\maketitle
\begin{abstract}
While concepts and tools from Theoretical Computer Science are regularly applied to,
and significantly support, software development for discrete problems,
Numerical Engineering largely employs recipes and methods 
whose correctness and efficiency is demonstrated empirically.
We advertise \emph{Real Complexity Theory}: 
a resource-oriented foundation 
to rigorous computations over continuous universes such as real numbers,
vectors, sequences, continuous functions, and Euclidean subsets:
in the bit-model by approximation up to given absolute error. 
It offers sound semantics (e.g. of comparisons/tests),
closure under composition, realistic runtime predictions, 
and proofs of algorithmic optimality by
relating to known classes like
$\calNP$, $\sharpP$, $\calPSPACE$.
\end{abstract}

%\begin{keywords}
%computational complexity, reliable numerics
%\end{keywords}

%\begin{AMS}
%65Y20, 68Q17
%\end{AMS}

%%%%%%%%%%%%%%%%%%%%%%%%%%%%%%%%%%%%%%%%%%%%%%%%%%%%%%%%%%%%%%%%
\noindent
Numerical methods permit digital computers, operating over sequences
of bits, to solve problems involving \emph{a priori} continuous objects such as
real numbers, functions, or operators.
Partial differential equations for instance are regularly
treated by discretizing the domain, thus approximating the
infinite-dimensional solution function space by
a high but finite dimensional one.
Other common (sub)problems include ordinary differential equations,
numerical integration and differentiation, or maximizing
some objective function subject to certain constraints.
We thus record that numerical science has devised a variety of
methods working impressively well in practice in terms
of an intuitive conception of efficiency.

Formal notions of algorithmic efficiency, on the other hand,
are at the core of Computer Science and have led to
the well-established complexity theory with famous
classes like $\calP$ and $\calNP$.
They capture the computational difficulty inherent to a
fully specified problem, rather than the cost 
of some method solving certain instances of it:
thus providing a sound framework for comparing any specific
algorithm against an (usually unknown) optimal one ---
in the discrete realm, that is, for problems
over integers or graphs encoded as finite sequences of bits.
Concerning real problems, we quote from \cite[p.412]{Linz}
and from \cite[\S1.4]{BCSS}:

\begin{quote}\it
How do engineers deal with the problem of assigning some measure of 
reliability to the numbers that the computer produces? Over the years, 
I have sat on many Ph.D. qualifying examinations or dissertation defenses 
for engineering students whose work involved a significant amount of 
numerical computing. In one form or another, I invariably ask two questions: 
``Why did you choose that particular algorithm?'' and 
``How do you know that your answers are as accurate as you claim?'' 
The first question is usually answered confidently, using such terms as 
``second-order convergence'' or ``von Neumann stability criterion''. 
The next question, alas, tends to be embarrassing. After an initial 
blank or hostile stare, I usually get an answer like 
``I tested the method with some simple examples and it worked'', 
``I repeated the computation with several values of $n$ and the 
results agreed to three decimal places'', or more lamely, 
``the answers looked like what I expected''. So far, I have not 
faulted any student for the unsatisfactory nature of such a response. 
One reason for my reluctance to criticize is that I have really 
nothing better to offer. \emph{Rigorous analysis is out of the question.}

\bigskip
 The developments described in the previous section
 (and the next) have given a firm foundation to computer
 science as a subject in its own right. Use of the Turing
 machines yields a unifying concept of the algorithm
 well formalized. [\ldots] The situation in numerical analysis
 is quite the opposite. Algorithms are primarily a means to
 solve practical problems. There is not even a formal definition
 of algorithm in the subject. [\ldots] Thus we view numerical
 analysis as an eclectic subject with weak
 foundations; this certainly in no way denies its great
 achievements through the centuries.
\end{quote}
However nowadays we do have a sound, realistic, and applicable 
theory of computations over continuous universes in the bit-cost model
approximations up to guaranteed absolute error $2^{-n}$:
Initiated by Alan Turing
in the very same work \cite{Turing37} that introduced the machine now named after him
--- and before he `invented' matrix condition numbers in 1948 ---
\emph{Recursive Analysis} has developed into a sound algorithmic foundation
to verified/reliable/rigorous (not necessarily interval) numerics 
and to computer-assisted proofs \cite{Rump,Plum} in unbounded precision;
cmp.~\cite{Siam100Digits,Braverman2,Braverman3}.
Traditionally focused on the computational contents of existence claims 
from classical analysis (see Items a,b,e,g,k,m in the below Example),
incomputability results often encode the Halting Problem
in a clever way; while positive ones usually devise
algorithms and establish their correctness.
Starting with \cite{KoFriedman}, however, the more refined view of complexity
has received increasing attention and in the last few years accumulated
a striking momentum \cite{Ko91,WeihrauchComplexity,SchroederComplexity,AkiSTOC}
with quantitative notions of efficiency 
and optimality of actual numerical computations!
Here are some of its main features:

\begin{itemize}
\item Including \emph{transcendental calculations}
 such as the exponential function:
 no restriction to algebraic numbers
\item \emph{Guaranteed output approximations}
 suitable (among others) for testing inner-mathematical conjectures
\item up to \emph{absolute error ${2^{-n}}$}
 roughly corresponding to $n$ valid binary digits after the radix point:
 renders addition of real numbers computable in time linear in $n$.
\item \emph{Adaptive precision} for intermediate calculations
 beyond the paradigmatic chains of hardware floats.
\item \emph{Fully specified algorithms}
 of guaranteed behavior on explicit classes of admissible inputs.
\item \emph{Parameterized analyses} asymptotically for $n\to\infty$
\item \emph{quantitatively} with respect to \emph{resources} like running time,
 memory (often the harder constraint), or \#processors/cores
\item \emph{Relating to standard complexity classes}
 from Theoretical Computer Science such as
 $\calLogspace\subseteq\calNC\subseteq\calP\subseteq\calNP\subseteq\calP^{\sharpP}
 \subseteq\calPSPACE\subseteq\calEXP$ etc.
\item Proving \emph{optimality} of an algorithm: for instance
 using adversary arguments in a bounded-precision adaptation of IBC
 or relative to complexity-theoretic conjectures like ``$\calP\neq\calNP$''.
\item Based on \emph{Turing machines} 
 for a formal foundation of numerical calculations in the bit-cost model that yield
\item both \emph{practical predictions} of the behaviour of actual implementations 
\item and \emph{closure under composition,} both of computable and polynomial-time
 computable functions: a prerequisite to \emph{modular software development}
 relying crucially on both real number outputs and inputs being given by approximations!
\item Actual programming in \emph{imperative object-oriented higher-level languages} 
 such as \texttt{C++} using libraries
 that implement a new data type \texttt{REAL} which through
 overloading support `exact' operations, thus facilitating
\item \emph{Rapid numerical prototyping} with accuracy/stability issues
 taken care of by the system transparently to the user.
\item Modified but \emph{consistent semantics} for tests (multivalued)
 and (partial) branching: because
 equality ``$x=0$?'' is provably not semi-decidable.
\item \emph{Interface declarations} of \texttt{C++} functions provided
 by the theory proving certain enrichment {\rm\cite{Advice}}
 of the continuous data necessary and sufficient
 for (say, polynomial-time) computability {\rm\cite{Promises}}.
\end{itemize}
\noindent
This flourishing field combines real (and complex)
analysis with theoretical computer science:

\begin{definition} \lab{d:CCA}
\begin{enumerate}
\item[a)]
Computing a real number $x$ means %(a Turing machine with one-way output tape)
printing some infinite sequence $a_n$ of integers (in binary without
leading zeros) as mantissae/numerators to dyadic rationals $a_n/2^{n}$
that approximate $x$ up to absolute error $1/2^n$. 
\item[b)]
More generally, a real sequence $(x_j)$ means producing an integer
double sequence $a_{j,m}$ with $|x_j-a_{j,m}/2^{m}|\leq2^{-m}$.
Formally, the elements of said sequence occur in order
according to the \emph{Cantor pairing function}
\begin{equation} \label{e:Pairing1}
 \IN\times\IN \ni (j,m) \;\mapsto\; \langle j,m\rangle \dfeq j+(j+m)\cdot(j+m+1)/2 \in \IN
\dfeq \{0,1,2,\ldots\} \end{equation}
that is, the output consists of the single integer sequence
$(a_n)$ with $n=\langle j,m \rangle$.
\item[c)]
Computing a univariate and possibly partial real function
$f \pcolon \IR\to\IR$ amounts to converting
every given sequence $(a_m)\subseteq\IZ$
with $|x-a_m/2^{m}|\leq1/2^m$ for any $x\in\dom(f)$
into a sequence $(b_n)\subseteq\IZ$
with $|f(x)-b_n/2^{n}|\leq1/2^n$.
\item[d)]
For some mapping $t\dffn\IN\to\IN_+\dfeq\{1,2,\ldots\}$,
the above computations are said to run in time $t(n)$
if the $n$-th integer output appears within at most $t(n)$ steps.
\emph{Polynomial time} (or \emph{polytime} for short)
means running time bounded by some polynomial $t\in\IN[X]$.
Similarly for exponential time and polynomial space.
\item[e)]
Generalizing (c) and (d),
consider a partial real multivalued function
$f \pcolon \IR^d\times\IN\toto\IR^e\times\IN$.
Computing $f$ amounts to converting,
for every $(\vec x,k)\in\dom(f)$,
given $k\in\IN$ and any sequence $(\vec a_m)\subseteq\IZ^d$
with $\|\vec x-\vec a_m/2^{m}\|\leq1/2^m$,
into $\ell\in\IN$ and some sequence $(\vec b_n)\subseteq\IZ^e$
with $\|\vec y-\vec b_n/2^{n}\|\leq1/2^n$ 
for some $(\vec y,\ell)\in f(\vec x,k)$.

Such a computation is said to run in \emph{fully polynomial time}
if $\vec b_n$ appears within a number of steps
at most polynomial in $n+k$, and $\ell$ is bounded
by some polynomial in $k$ only.
\end{enumerate}
\end{definition}
Parameterized complexity theory as in (e) relaxes
Condition~(d) requiring the running time to be bounded 
in terms of the output precision only; see Example~\ref{x:Main}(f) and (s) below.
Fully polynomial-time computable functions are closed under composition.

\begin{myexample} \label{x:Main} 
\setlength\leftmargini{0.17in}
\begin{description}
\item[\rm a) Monotone Convergence Theorem:]
  There exists a computable, monotonically increasing sequence $(x_j)\subseteq[0;1]$
  with incomputable limit $\sup_j x_j$, that is,
  $(x_j)$ admits no recursively bounded rate of convergence {\rm\cite{Specker49}}.
\item[\rm b) Bolzano--Weierstra\ss:]
  There exists a computable sequence $(x_j)\subseteq[0;1]$ such that
  no accumulation point can be computed even relative
  to the Halting Problem as oracle \mycite{Theorem~3.6}{noCH}.
\item[\rm c) Real-Closed Fields:]
  Sum, product, and inverse of (polynomial-time) computable reals are again (polynomial-time) computable. 
  For every complex polynomial with computable coefficients,
  all its real roots are again computable {\rm\cite{Specker69}}.
  In fact, if the polynomial's coefficients are polynomial-time computable,
  then so are its real roots {\rm\cite{Schoenhage}}.
\item[\rm d) Matrix Diagonalization:]
  Every real symmetric $d\times d$--matrix $M$ with computable entries
  admits a computable basis of eigenvectors.
  However such a basis cannot in general be computed from approximations to $M$ only;
  whereas restricted to non-degenerate $M$ 
  and, more generally, when providing in addition to approximations to $M$ the integer
  $\Card\sigma(M)\in\{1,\ldots,d\}$ of distinct eigenvalues, it can \mycite{\S3.5}{LA}
  -- and this amount of so-called \emph{discrete enrichment} is optimal {\rm\cite{Advice}}.
\item[\rm e) Power Series:]
  There exists a computable real sequence $\bar c=(c_j)$ whose
  radius of convergence $R(\bar c):=1/\limsup_j |c_j|^{1/j}$ is 
  not computable even relative to the Halting Problem as oracle \mycite{Theorem~6.2}{ArithHierarchy}.
  Moreover power series evaluation 
  $x\mapsto\sum_j c_j x^j$ is in general not computable on $(-R,R)$ \mycite{Example~6.5.1}{Weihrauch};
  whereas, for every fixed $0<r<R$, it is computable on $[-r;r]$ \mycite{Theorem~4.3.12}{Weihrauch}.
\item[\rm f) Some Polynomial-Time Computable Functions:]
  Addition $(x,y)\mapsto x+y$ and multiplication $(x,y)\mapsto x\cdot y$
  on the real interval $[-2^k; 2^k]$ are computable in time polynomial in $k+n$,
  where $n$ denotes the output precision in the sense of guaranteed approximation
  up to absolute error $1/2^n$.
  The exponential function (family) on $[-2^k; 2^k]$ is computable in time
  polynomial in $2^k+n$ but not in time depending only on $n$.
  Reciprocals $[2^{-k};\infty)\ni x\mapsto 1/x$ are computable in time polynomial in $k+n$.
  The square root is computable on $[0;2^k]$ in time polynomial in $k+n$.
The following explicit function is computable in exponential
time, but not in polynomial time --- even relative to any oracle:
\begin{equation} \label{e:Exp}
 \hexp \dffn [0; 1] \to \IR,  \qquad
 \hexp(0) \dfeq 0, \quad
 0 < x \mapsto \hexp(x) \dfeq 1/{\ln(e/x)} \enspace .
\end{equation}
\item[\rm g) Maxima:]
  For every computable $f \dffn [0;1]\to\IR$, the parametric maximum function
  $\MAX(f):[0;1]\ni x\mapsto \max\{f(t):0\leq 1\leq x\}$ is again computable.
  However there exists a computable smooth $f$ which
  attains its maximum at (many, but) no computable $x$ {\rm\cite{Specker59}}.
\item[\rm h) Integration:] For every computable $f \dffn [0;1]\to\IR$,
  the indefinite Riemann integral $\int f:[0;1]\ni x \mapsto \int_0^x f(t)\,dt$ is again computable.
\item[\rm j) Derivatives:] There exists a computable $f\in C^1[0;1]$ 
  with $f'$ non-computable {\rm\cite{Myhill}}. \\
%  On the other hand 
Every computable $f\in C^2[0;1]$
  has a computable derivative \mycite{Corollary~6.4.8}{Weihrauch}.
\item[\rm k) Ordinary Differential Equations:] There exists a computable (and thus continuous)
  $f \dffn [0;1]\times[-1;1]\to[-1;1]$ such that the initial value problem
\begin{equation} \label{e:IVP}
  \dot{y}(t) \;=\; f\big(t,y(t)\big), \quad y(0)=0
\end{equation}
  has (many, but) no computable solution $y \dffn [0;1]\to[-1;1]$ {\rm\cite{PER79}}.
  However if $f$ is computable \emph{and Lipschitz},
  then the (now unique) solution $y$ is again computable.
\item[\rm m) Partial Differential Equations:] There exists a computable $f\in C^1(\IR^3)$
  such that the strong solution $u=u(t,x,y,z)$ to the
  linear Wave Equation in 3D
\begin{equation} \label{e:Wave}
  \tfrac{\partial^2}{\partial t^2} u \;=\;
  \tfrac{\partial^2}{\partial x^2} u \;+\;
  \tfrac{\partial^2}{\partial y^2} u \;+\;
  \tfrac{\partial^2}{\partial z^2} u , \quad
    u(0,\cdot)=f, \quad \tfrac{\partial}{\partial t} u(0,\cdot)=0
\end{equation}
  is incomputable at $(1,0,0,0)$ {\rm\cite{PER81}}. 
  Its Sobolev solution however is computable {\rm\cite{ZhongWave2}}.
\item[\rm n) Euclidean subsets:]
  There exist compact subsets $A,B\subseteq\IR$ with computable
  distance function whose intersection $A\cap B\neq\emptyset$
  has a non-computable distance function \mycite{Exercise~5.1.15}{Weihrauch}.
  There exists a regular compact subset of the plane
  whose \emph{weak membership oracle} \cite{GLS} is polynomial-time computable,
  while its distance function is so if and only if $\calP=\calNP$ holds 
  {\rm\cite{Braverman0}}. 
  For any convex compact Euclidean subset on the other hand,
  the computational complexity of its distance function
  and its \emph{weak membership oracle} are parameterized
  polynomially related {\rm\cite{Roesnick}}.
\item[\rm o) Maximum values, again:]
  Whenever $f \dffn [0;1]\to\IR$ is polynomial-time computable,
  $\MAX(f)$ is computable in exponential time and polynomial space.
  In case $\calP=\calNP$, $\MAX(f)$ is even polynomial-time computable.
  Conversely, there exists a polyno\-mi\-al-time computable 
  $f \in C^\infty[0;1]$ such that polynomial-time computability of $\MAX(f)$ 
  implies  $\calP = \calNP$ \mycite{Theorem~3.7}{Ko91}.
  For analytic functions $f$, however, $\MAX(f)$ is polynomial-time computable
  whenever $f$ is {\rm\cite{Mueller87}}.
\item[\rm p) Integration, again:]
  Whenever $f \dffn [0;1]\to\IR$ is polynomial-time computable, 
  $\int f$ is computable in exponential time and polynomial space. 
  In case $\calFP = \sharpP$, $\int f$ is even computable in polynomial time.
  Conversely, there exists a polynomial-time computable 
  $f\in C^\infty[0;1]$ such that polynomial-time computability of $\int f$ 
  implies $\calFP = \sharpP$ \mycite{Theorem~5.33}{Ko91}.
  For analytic functions $f$, however, $\int f$ is polynomial-time computable
  whenever $f$ is {\rm\cite{Mueller87}}.
\item[\rm q) ODEs, again:]
  Whenever $f \dffn [0;1]\times[-1;1]\to[-1;1]$ is polynomial-time computable
  and Lipschitz continuous, then the unique solution $y$ to
  Equation~(\ref{e:IVP}) is computable in exponential time and polynomial space.
  In case $\calP=\calPSPACE$, $y$ is even computable in polynomial-time.
  Conversely, there exists a polynomial-time computable $f\in C^1$ such that
  polynomial-time computability of solution $y$ 
  implies $\calP = \calPSPACE$ {\rm\cite{AkiODE,Ota}}.
  If analytic $f$ is polynomial-time computable,
  then so is $y$ {\rm\cite{Mueller95}}.
\item[\rm r) PDEs, again:]
  Recall the Dirichlet Problem for 
  Poisson's linear partial differential equation:
\begin{equation} \label{e:Poisson}
  f \;=\; \Delta u \quad\text{on }\Omega, \qquad
    u=g \quad\text{on }\partial\Omega \enspace.
\end{equation}
  On the Euclidean unit ball $\Omega\subseteq\IR^d$
  for polynomial-time computable $f:\Omega\to\IR$ and $g:\partial\Omega\to\IR$, 
  the strong solution $u$
  exists and is computable in exponential time and polynomial space.
  In case $\calFP = \sharpP$, it is even computable in polynomial time.
  Conversely, there exist polynomial-time computable smooth
  $f,g$ such that polynomial-time computability of $u$
  implies $\calFP = \sharpP$ {\rm\cite{Poisson}}.
\item[\rm s) Analytic and Gevrey Functions:]
  For $\gamma,B,\ell\in\IN$ let 
  $\Gevrey^\gamma_{B,\ell}[0;1]$ denote the class of all
  infinitely differentiable functions $f:[0;1]\to\IR$
  satisfying $|f^{(j)}(x)|\;\leq\; B\cdot \ell^j\cdot j^{j\gamma}$.
  Then $\bigcup_{B,\ell}\Gevrey^1_{B,\ell}[0;1]$ coincides with the
  class of real analytic functions. Moreover (i) evaluation,
  (ii) addition, (iii) multiplication, (iv) differentiation,
  (v) integration, and (vi) maximization on $\Gevrey^\gamma_{B,\ell}[0;1]$
  is uniformly computable in time polynomial in $(n+\ell+\log B)^\gamma$;
  and this is asymptotically best possible {\rm\cite{Gevrey}}.
\end{description}
\end{myexample}
Items~(n) to (r) suggest numerical approaches to famous open problems
in discrete complexity theory. From a different perspective, they
demonstrate exponential-time behaviour of algorithms for several 
numerical problems to be optimal subject to said conjectures \cite{Ota2}.
Item~(s) `explains' for the phase transition between analytic to smooth
functions indicated in Items~(o) and (p). It also exhibits parameterized 
preductions to the actual behaviour of fully-specified algorithms --- as 
opposed to `methods' or `recipes' for a vaguely-defined task such as 
in the following quote from the \NAG library's documentation:

\begin{quote} \it
{\tt nag\_opt\_one\_var\_deriv(e04bbc)} \underline{normally} computes a
sequence of $x$ values which \underline{tend in the limit} to a minimum of
$F(x)$ subject to the given bounds.
\end{quote}
We particularly promote three concepts from logic
that have turned out essential in \emph{Real Complexity Theory}
with consequences for computational practice:

{\setlength\leftmargini{0.15in}
\begin{description}
\item[\it Nonextensional/multivalued Functions:]
  A (total) relation $f\subseteq X\times Y$ can alternatively be
  regarded as a partial set-valued mapping $f \dffn \subseteq X\to 2^Y\setminus\{\emptyset\}$,
  sometimes also written as $f \dffn \dom(f)\subseteq X\toto Y$
  with $\dom(f):=\{x\in X:\exists y\in Y: (x,y)\in f\}$,
  via $x\mapsto \{y\in Y:(x,y)\in f\}$.
  It corresponds to a computational semantics where an algorithm is
  permitted, given $x$ encoded in one way, to output some $y\in f(x)$
  but some possibly different $y'\in f(x)$ when given the same $x$
  encoded in another way. Such \emph{non-extensionality}
  models computational search problems well-known in the discrete case.
  However in the real setting this effect emerges naturally also 
  in subproblems even when computing only single-valued functions;
  see \mycite{\S2.3.6}{VascoDiss},
  \mycite{Exercise~5.1.13}{Weihrauch}, or {\rm\cite{Luckhardt}}.
\item[\it Discrete Enrichment:]
  Many incomputable (single or multivalued) problems $f \dffn X\toto Y$
  do become computable when providing, in addition to approximations
  to the arguments $x\in X$ certain additional information,
  a concept well-known as \emph{enrichment} in logic {\rm\cite[p.238/239]{KreiselMacIntyre}}.
  In many practical cases it suffices to enrich the given $x$
  with some suitable integer $k=k(x)$,
  usually even from a bounded range. For instance
  according to Example~\ref{x:Main}(d),
  only the second line in the following \texttt{C++} fragment
can lead code correctly
returning some eigenvector to a given real symmetric $2\times2$--matrix:
\begin{verbatim}
void EV(REAL A11, REAL A12, REAL A22,                 REAL &EVx, REAL &EVy)
void EV(REAL A11, REAL A12, REAL A22, int degenerate, REAL &EVx, REAL &EVy)
\end{verbatim}
\item[\it Parameterized Complexity:]
  Theoretical Computer Science traditionally considers
  the worst-case resource consumption in dependence on the binary
  input length parameter $n$. In the real realm
  inputs are infinite, and $n$ instead denotes the output precision.
  However in both disciplines additional parameters $\vec k$
  allow for a finer-grained analysis and more realistic predictions;
  recall Definition~\ref{d:CCA}(e) with Example~\ref{x:Main}(f) and (s).
  In the discrete case this leads to the field of
  \emph{Parameterized Complexity}. In the real case,
  condition numbers are (but) one example of a secondary parameter;
  the complexity analyses of addition and exponential
  function from the above examples demonstrate also other
  natural choices. Discrete enrichment often
  simultaneously serves as a secondary complexity parameter;
  cmp. Example~\ref{x:Main}(s).
\end{description}}

\noindent
To conclude, Real Complexity Theory provides a computer-scientific
foundation to Numerics bridging from Recursive Analysis to practice. 
It asserts in a sound setting that common problems with guaranteed 
precision are surprisingly hard in the worst case, even restricted
to smooth functions. On the other hand recipes and methods do work
surprisingly well in practice. This gives raise to one 
among many questions and challenges for future research:

\begin{challenge} \lab{c:Main}
\begin{enumerate}
\item[a)] 
  Formally capture the class of `practical' functions that renders
  standard operations (i) to (vi) from Example~\ref{x:Main}(s)
  polynomial-time computable.
\item[b)]
  Devise, similarly to Example~\ref{x:Main}(s), 
  a parameterized complexity theory 
  and reliable implementation of ODE solutions.
\item[c)] 
  Identify a suitable notion of resource-bounded computation
  on Sobolev spaces, and characterizes the complexity
  inherent to the last line of Example~\ref{x:Main}(m).
\item[d)]
  Develop a sound computability and complexity-theoretic
  foundation of recent approaches in numerical engineering
  to shape and topology optimization.
\end{enumerate}
\end{challenge}
We close with some programming examples evolving around
iterating the \emph{Logistic Map}:
\begin{equation} \label{e:Logistic}
[0;1]\;\ni\; x\;\mapsto\; r\cdot x\cdot (1-x)\;\in\;[0;1], \qquad 1<r<4 \enspace .
\end{equation}
It is well-known to exhibit chaotic behaviour for many values of the parameter
$r$ beyond $3.56995$, with the exception of isolated so-called
\emph{islands of stability} for example at $r=1+\sqrt{8}\approx 3.82843$.
For $r:=15/4=3.75$ and $x_0:=1/2$ and $m=30,40,85,100,200,500,1000,10\,000$,
the reader is encouraged to actually run the code fragments below,
and to compare the results they produce: in 
\texttt{Matlab} (left box), \texttt{Maple} (middle), and in
\texttt{C++} (right box)
with first line alternatingly replaced by \texttt{\#define REAL double},
by \texttt{\#define REAL long double}, and by \texttt{\#include "iRRAM.h"}.
The latter refers to the \iRRAM{} library 
freely available 
from \myurl{http://www.informatik.uni-trier.de/iRRAM/}

\medskip
\medskip
\medskip
\noindent
\begin{minipage}{0.32\textwidth}
\begin{Verbatim}[frame=single]
function y=logistic(m) 
x=1/2; r=vpa(15/4);
for j=1:m x=r*x*(1-x);
end; y=x; end
\end{Verbatim}
\end{minipage}\hfill
\begin{minipage}{0.25\textwidth}
\begin{Verbatim}[frame=single]
logistic:=proc(m)
local j,x,r;
x:=1/2; r:=15/4;
for j from 1 to m 
do x:=r*x*(1-x) 
end do end proc;
\end{Verbatim}
\end{minipage}\hfill
\begin{minipage}{0.36\textwidth}
\begin{Verbatim}[frame=single]
#define REAL float
REAL logistic(int m) { 
REAL x = REAL(1)/REAL(2), 
    r = REAL(15)/REAL(4);
while (m--) x = r*x*(1-x); 
return(x); }
\end{Verbatim}
\end{minipage}
%

%%%%%%%%%%%%%%%%%%%%%%%%%%%%%%%%%%%%%%%%%%%%%%%%%%%%%%%%%%%%%%%%%%%%%%%%%%

\end{document}